\documentclass{ws-procs9x6}

\begin{document}

\title{{\em PulsarSpectrum}: simulating gamma-ray pulsars for the GLAST mission}
\author{Massimiliano Razzano, Luca Latronico, Nicola Omodei, Gloria Spandre}

\address{Istituto Nazionale di Fisica Nucleare, sezione di Pisa,\\
 Largo B.Pontecorvo 3, 56100 Pisa\\
  E-mail: massimiliano.razzano@pi.infn.it}

\maketitle

\abstracts{We present here an overview of {\em PulsarSpectrum}, a
program that simulates the gamma ray emission from pulsars. This
simulator reproduces not only the basic features of the observed
gamma ray pulsars, but it can also simulate more detailed effects
related to pulsar timing. It is a very useful tool to understand the
GLAST capabilities in the pulsar science. }

\section{Introduction}
Pulsars are among the most exciting gamma ray sources in the
Universe and can serve as unique sites for the study of emission
processes in extreme physical environments. The Gamma ray Large Area
Space Telescope (GLAST) will increase dramatically our knowledge of
gamma ray pulsars physics. In particular the Large Area Telescope
(LAT), the main GLAST instrument, will provide more detailed
observations of the known gamma ray pulsars and potentially will
discover many new pulsars that emit gamma rays. To better understand
the capabilities of GLAST for pulsar science we developed {\em
PulsarSpectrum}, a program that simulates gamma ray emission from
pulsars. This simulator can be easily interfaced with the Monte
Carlo software that simulates the response of the LAT.

\subsection{Gamma ray pulsars in the GLAST era}
Pulsars have been associated with high energy gamma ray astronomy
since the first experiment capable to resolve point sources. The
{\em Small Astronomy Satellite}(SAS-2), launched in 1972, identified
 three intense sources in the sky, that in the following turned out
to be the Vela, Crab and Geminga pulsars. A signal modulated at the
radio period was found for Vela [\refcite{Kniff74}] and Crab
[\refcite{Thomp75}], but for Geminga no radio counterpart was found
and this source remained quite mysterious until it was identified as
an X-ray and gamma-ray pulsar. The {\em Energetic Gamma Ray
Experiment} (EGRET) aboard the {\em Compton Gamma Ray Observatory}
(CGRO) detected three more pulsars, B1706-44 [\refcite{Thomp92}],
B1055-52 [\refcite{Fierro93}] and later B1951+32
[\refcite{Raman95}]. Another one, B1509-58 was observed at energies
between 60 keV and 2 MeV by the three CGRO experiments, the {\em
Burst and Transient Source Experiment} (BATSE)[\refcite{Wilson93}],
the {\em Oriented Scintillation Spectrometer Experiment} (OSSE),
[\refcite{Ulmer93}] and the {\em Imaging Compton Telescope}
(COMPTEL)[\refcite{Bennet93}], but it was not seen by EGRET, because
the spectrum shows a cutoff at energies lower than the instrument's
threshold. Recently the millisecond pulsar PSR 0218+4232 was
detected by analyzing the EGRET data [\refcite{Kuiper00}],
increasing to eight the number of detected gamma ray pulsars, seven
of them are radio-loud pulsars and only Geminga seems to be radio
quiet.

\begin{figure}[ht]
\epsfxsize=6cm   
\centerline{\epsfxsize=5.5cm\epsfbox{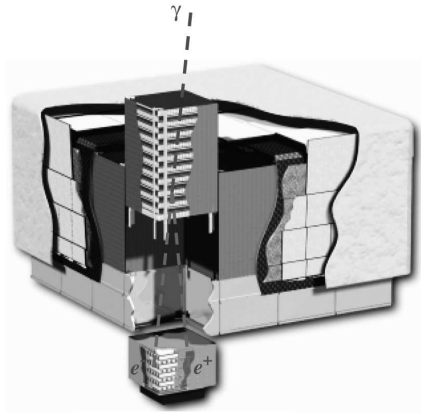}} \caption{The
GLAST Large Area Space Telescope. An incident gamma ray converts
into a $e^{-}e^{+} $ pair\label{lat}}
\end{figure}

The GLAST Large Area Telescope (LAT) (fig \ref{lat}), is a pair
conversion telescope based on the most advanced high energy
detectors. It consists in a precision silicon strip tracker, a
hodoscopic calorimeter and a segmented anticoincidence shield for
particle background rejection. The LAT high sensitivity (2 10$^{-9}$
ph/cm$^{2}$/s) and effective area ($>$8000 cm$^{2}$) will permit the
discovery of a lot of new pulsars: the estimates range between tens
to hundreds depending upon the model adopted. Moreover the low dead
time of the detector (20 $\mu$s) will allow the detailed
reconstruction of pulsar lightcurves. One of the most exciting
possibilities of the LAT will be the coverage of the energy window
from 30 GeV up to 300 GeV, a still unexplored range. At these
energies the theoretical models make different previsions on the
high energy spectral cutoff and the spectral coverage of LAT will be
of primary importance for constraining and discriminating among the
models. In order to study the LAT response to specific gamma ray
sources, various simulation packages have been developed. Here we
present {\em PulsarSpectrum}, that can simulate the observed pulsars
and create new fake pulsars for the LAT threshold detection
identification.

\section{The {\em PulsarSpectrum} simulator}

\subsection{Overview of the simulator}
The basic idea behind {\em PulsarSpectrum} is to construct a
2-dimensional histogram representing the differential flux vs.
energy and pulsar phase. This histogram contains all the basic
informations about lightcurve and spectrum. How it is built depends
upon the model we want to use: a phenomenological model, based only
on observations, or a physical one, based on a specific theoretical
model. At present only a phenomenological model has been implemented
because it is more flexible and completely independent from the
chosen theoretical scenario. The input parameters of the simulator
can be divided in two categories:
\begin{itemize}
\item {\em Observational parameters}, (i.e the flux or the ephemerides);
\item {\em Model-dependent parameters}, (i.e. the spectral index);
\end{itemize}
These parameters are placed in two specific data files used by both
{\em PulsarSpectrum} and the LAT simulation tools. {\em
PulsarSpectrum} creates the lightcurve and the spectrum from these
parameters and combines them to obtain a two-dimensional matrix that
represents the flux in ph/m$^2$/s/keV. The photons are then
extracted such that the interval between two subsequent photons is
determined by the flux integrated over the energy range of interest.

\subsection{The phenomenological model}
The currently implemented phenomenological model allows the user to
generate pulsar lightcurves in a general way using a single or
double Lorenzian peak profile whose shape is determined from random
generated numbers. The lightcurve can be generated alternatively
from a user-provided profile. This is useful for simulating the
already observed gamma ray pulsars. The spectral shape is assumed to
be a power law with exponential cutoff (according to
[\refcite{NelDj95}]), as in the observed gamma-ray pulsars and can
then be modeled as:
\begin{equation}
\frac{dN}{dE} =  K(\frac{E}{E_{n}})^aexp(\frac{E}{E_0})^{-b}
\end{equation}
The normalisation constant  K is determined by the photon flux above
100 MeV and the other four parameters can be varied; the values for
the EGRET pulsars are obtained from real data by fitting procedures
(e.g. [\refcite{NelDj95}]).
\begin{figure}[ht]
\centerline{\epsfxsize=7cm\epsfbox{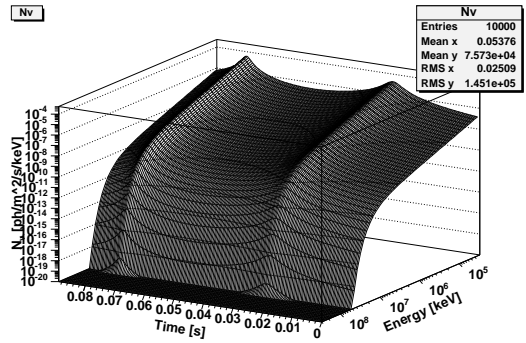}} \caption{An
example of 2-dimensional histogram created by PulsarSpectrum.
\label{nv}}
\end{figure}
\subsection{Timing issues}
Once the differential flux histogram is created the time interval
between two subsequent photons is computed according to the flux. If
the previous photon came at time t$_{0}$ the next photon will appear
at $\tilde {t}$ such that:
\begin{equation}
{A_{eff}}\int_{t_0}^{\tilde t}
\int_{E_1}^{E_2}\frac{dN}{dEdAdt}\,dEdAdt\ = 1
\end{equation}
The interval between two photons is computed assuming that the
pulsar period does not change with time and the photons arrival
times are computed into a reference system fixed relative to stars.
This is not the "real world". Pulsar timing is affected by more
complicate effects, as  (1)- The motion of the spacecraft through
the Solar System and the relativistic effects due to gravitational
well of the Sun (see \ref{barydecorr}); (2)- Period changes with
time (see \ref{pchange}). For pulsars in binary systems an
additional modulation to the orbital period should be taken into
account. For a precise pulsar simulator intent to produce a
realistic list of photon arrival times we need to include all these
effects (to transform to the observational frame). All these
procedures are now implemented in the code and only the binary
demodulation is not yet implemented. The real arrival time of a
photon from a pulsar must be first barycentered and then phase
assigned.
\subsubsection{Barycentric effects}\label{barydecorr}
The first step to analyze pulsar data is the conversion from the
arrival times at the spacecraft, usually expressed in Terrestrial
Time TT or TAI, to the arrival times at the Solar System barycenter,
expressed in Barycentric Dynamical Time TDB. Taking into account
both the motion of spacecraft through space and the general
relativistic effects due to the gravitational field of the Sun (i.e.
Shapiro delay), the simulator computes the opposite of the
barycentric correction by considering the position of the Earth and
of the spacecraft in the Solar System, and the position of the Sun.
The accuracy for the computation of these "de-corrections" is
hard-coded in the program.
\subsubsection{Period change and ephemerides}\label{pchange}
The rotational energy of a radio pulsar decreases with time and
hence the period increases with time. For gamma ray pulsar science
the radio ephemerides are fundamental for assigning the correct
phase to each photon. If we know the frequency {\em f$(t_{0}$)} and
its derivatives {\em $\dot{f} (t_{0})$} and {\em $\ddot{f} (t_{0})$}
at a certain time t$_{0}$, known as {\em epoch}, the phase is then:
\begin{equation}\label{phit}
\phi(t) = int[ f(t_{0})(t-t_{0}) + \frac{1}{2}\dot{f}
(t_{0})(t-t_{0})^{2} + \frac{1}{6}\ddot{f} (t_{0})(t-t_{0})^{3}].
\end{equation}
The interval between two photons must be "de-corrected" for this
effect. In the parameters file the user can specify a set of
ephemerides with the relative epoch of validity expressed in
Modified Julian Date. The simulator then computes the opportune
arrival time such that, after applying the barycentric corrections
and the equation \ref{phit}, the correct phase is obtained.

\begin{figure}[ht]
\centerline{\epsfxsize=8cm\epsfbox{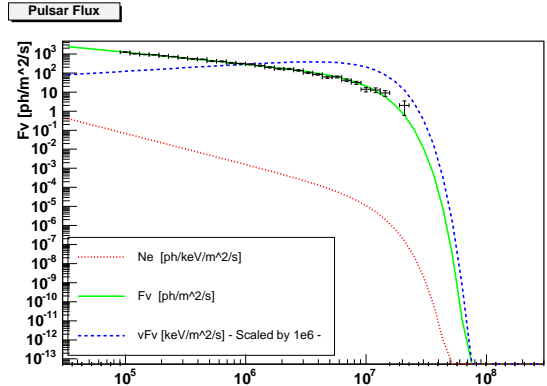}} \caption{An
example of the spectrum of a simulated pulsar created by
PulsarSpectrum. The point represents the extracted photons of energy
above 100 MeV. \label{fv}}
\end{figure}

\section{Conclusions}
{\em PulsarSpectrum} has been successfully used by the GLAST
collaboration for testing the functionality of the LAT Science
Analysis Tools, a set of analysis programs specifically designed to
analyse the LAT data after launch. Periodically these tools are
evaluated through specific Checkout phases. In absence of real data
there is strong need of detailed simulated data, most of them
provided by {\em PulsarSpectrum}. {\em PulsarSpectrum} is the most
probable candidate to be used as pulsar simulator in the GLAST Data
Challenge 2, an analysis phase where some months of LAT data are
simulated. A new interesting opportunity is now raising from the
creation of the LAT Science Groups, to study GLAST science
opportunities on specific topics. Our simulator has all the
characteristics to fit well the requirements of the LAT Pulsar
Science Group.

\end{document}